# Hybrid integration of deterministic quantum dots-based single-photon sources with CMOS-compatible silicon carbide photonics

Yifan Zhu[1,4], Wenqi Wei[2], Ailun Yi[1], Tingting Jin[1,4], Chen Shen[1,3], Xudong Wang[1,4], Liping Zhou[1,4], Chengli Wang[1,4], Weiwen Ou[1], Sannian Song[1,4], Ting Wang[2,4], Jianjun Zhang[2,4], Xin Ou[1,4,*] and Jiaxiang Zhang[1,4,*]

1, State Key Laboratory of Functional Materials for Informatics, Shanghai Institute of Microsystem and Information Technology, Chinese Academy of Sciences, Shanghai 200092, China

2, Beijing National Laboratory for Condensed Matter Physics, Institute of Physics, Chinese Academy of Sciences, Beijing 100190, China

3, Institute for Quantum Science and Engineering and Department of Physics, Southern University of Science and Technology, Shenzhen 518055, China

4, Center of Materials Science and Optoelectronics Engineering, University of Chinese Academy of Sciences, Beijing 100049, China

*Corresponding authors: ouxin@mail.sim.ac.cn, jiaxiang.zhang@mail.sim.ac.cn

Yifan Zhu, Wenqi Wei, Ailun Yi contributed equally to the work.

**Abstract**: Thin film 4H-silicon carbide (4H-SiC) is emerging as a contender for realizing large-scale optical quantum circuits due to its high CMOS technology compatibility and large optical nonlinearities. Though, challenges remain in producing wafer-scale 4H-SiC thin film on insulator (4H-SiCOI) for dense integration of photonic circuits, and in efficient coupling of deterministic quantum emitters that are essential for scalable quantum photonics. Here we demonstrate hybrid integration of self-assembled InGaAs quantum dots (QDs) based single-photon sources (SPSs) with wafer-scale 4H-SiC photonic chips prepared by ion slicing technique. By designing a bilayer vertical coupler, we realize generation and highly efficient routing of single-photon emission in the hybrid quantum photonic chip. Furthermore, we realize a chip-integrated beamsplitter operation for triggered single photons through fabricating a 1×2 multi-mode interferometer (MMI) with a symmetric power splitting ratio of 50:50. The successful demonstration of heterogeneously integrating QDs-based SPSs on 4H-SiC photonic chip prepared by ion slicing technique constitutes an important step toward CMOS-compatible, fast reconfigurable quantum photonic circuits with deterministic SPSs.

**Keywords**: silicon carbide photonics, hybrid integration, self-assembled quantum dots, single-photon sources, multimode interference beamsplitter

Integrated quantum photonic circuits have long been sought after due to their high degree of compactness, robustness, and stability for quantum photonic quantum information science[1-5]. They have been extensively explored as a resource of novel methods to improve the complexity and scalability of photonic quantum technologies, including quantum computing[6-7], quantum simulations[8-9], and quantum communications[10-11]. Currently, silicon is the core material for integrated quantum photonics. However, due to the absence of direct band gap and intrinsic electro-optic effect, silicon photonics is limited by the large losses from the use of off-chip quantum light sources and the carrier-injected plasma light modulation process[11-13]. Towards the development of advanced integrated photonics with ultrafast light modulators and chip-integrated quantum emitters, exploring novel materials with direct band gap and high electro-optic coefficient is highly demanded.

Silicon carbide (SiC), a wide band gap semiconductor, has recently emerged as a promising material platform for next-generation integrated photonics[14-15]. In addition to its high CMOS-compatible material property, SiC has established many benefits for integrated photonics owing to its exceptional optical properties, including the large second- and third-order nonlinearities[16-17], the wide band gap of 2.4~3.2eV[18], high refractive index (~2.6)[19-20]. Among 200 polytypes of SiC that are currently investigated, the cubic[21-22] and hexagonal polytypes[14, 23], corresponding to 3C- and 4H-SiC, are the most suitable materials for integrated photonics resulting from their stable form and well-established growth method. In the past decade, tremendous efforts have been devoted to develop photonic devices by using 3C-SiC material platform[21, 24-27]. However, 3C-SiC is a polycrystalline material and it suffers from large intrinsic material absorption loss due to the high density of material imperfections occurred in its heteroepitaxial growth[28]. In contrast, the hexagonal 4H-SiC can maintain the crystalline phase when processed into thin films, and it has much lower intrinsic loss than that of 3C-SiC. The recent work of realizing high-quality single-crystalline 4H-SiC thin film on insulator (4H-SiCOI) represents an important progress toward the development of SiC based integrated photonic circuits with ultra-low loss[15, 23]. Nevertheless, the preparation of 4H-SiC thin films reported in previous works requires sophisticated grinding and thinning techniques, and this complexity renders fabrication of wafer-scale 4H-SiCOI inconvenient[15]. Compared to the mechanical thinning technique, applying ion slicing technique to prepare 4H-SiCOI is practically favorable[29]. Considerable benefits are the mass production of wafer-scale 4H-SiCOI with a well-controlled thickness of the thin film and the industry-compatible device fabrication method, crucial for the achievement of large-scale integrated photonic circuits with high reproducibility and low costs[30-32].

A second challenge to leverage 4H-SiC for integrated quantum photonic applications is to create efficient single-photon sources (SPSs) on chip. Although 4H-SiC harbors a rich assortment of optically addressable spin defects, spatially resolvable single defects have proven extremely difficult due to the high dose of ion implantation used in the slicing process[33-35]. To circumvent this complication, heterogeneous integration of solid-state quantum emitters in a hybrid fashion is particularly appealing as it provides a viable route to combine different photonic building blocks that are well-developed on distinct material platforms in a single chip unit[36-40]. Of various quantum emitters reported so far, self-assembled quantum dots (QDs) contribute the most versatile systems because of their proven potential to deterministically emit single photons with high brightness, high purity, and indistinguishability[41-42]. In particular, they can be easily incorporated into planar photonic cavities so that near transform-limited single-photon emission can be realized. Up to now, great success has been achieved in integrating QDs-based SPSs on a variety of photonic platforms, with some examples being silicon[37, 40, 43-44], silicon nitride[45-46], aluminum nitride[47-49] and lithium niobate[36]. As compared to these photonic material platforms, 4H-SiC offers unique potential for on-chip quantum photonics, as it is compatible with CMOS nanofabrication and has a strong second-order optical nonlinearity. Successful integration of QDs-based SPSs with this novel material platform could open additional design space to integrated photonics and enable novel functionalities on chip that are not available on other material platforms. Despite the envisaged prospects, hybrid integration of self-assembled QDs with 4H-SiCOI material has remained unexplored yet.

In this letter, we demonstrate the first hybrid integration of deterministic SPSs with low-loss 4H-SiCOI photonic platform that is prepared by ion slicing technique. The quantum emitters are InGaAs QDs embedded inside GaAs nanophotonic waveguides. A pick-and-place technique is utilized to transfer GaAs nanophotonic waveguides onto 4H-SiCOI photonic chips, and a bilayer vertical coupler consisting of a tapered waveguide is employed to achieve a near unity coupling efficiency of QDs emission into the 4H-SiC waveguide. Furthermore, we deterministically couple photon emission from a single QD to a 50/50 multi-mode interferometer (MMI), and demonstrate an on-chip beamsplitter operation with triggered single photons. The hybrid quantum photonic chip based on self-assembled QDs and 4H-SiCOI material demonstrated in this work provide a viable routine towards industry-standard and CMOS-compatible integrated quantum photonic circuits with possible fast electro-optic light modulators and deterministic SPSs.

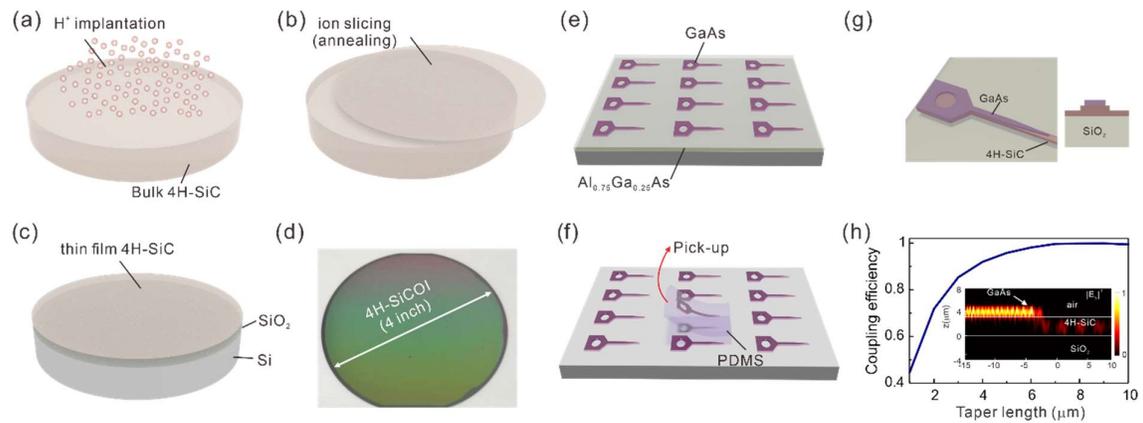

**Figure 1:** Ion slicing technique for preparing wafer-scale 4H-SiCOI material platform. The process includes a) hydrogen ion implantation to a bulk 4H-SiC wafer, b) thermal annealing to detach the 4H-SiC thin-film from the virgin substrate, c) direct wafer bonding to transfer the thin-film 4H-SiC onto a SiO$_2$/Si substrate, forming a heterogenous integrated 4H-SiCOI material platform. d) A typical microscopic image of the fabrication 4-inch 4H-SiCOI wafer. e) Nanofabrication of photonic waveguides on a 180 nm thick GaAs nanomembrane grown on a 200 nm thick Al$_{0.75}$Ga$_{0.25}$As sacrificial layer. e) Chemical wet etching was used to undercut the sacrificial layer in order to make the GaAs nanophotonic waveguides free-standing. The free-standing GaAs nanophotonic waveguide is then picked-up and deterministically transferred onto a 4H-SiCOI photonic chip by using a PDMS stamp. g) Sketch of a bilayer vertical coupler where GaAs nanophotonic waveguide is transferred onto a 4H-SiC ridge waveguide. h) Simulated coupling efficiency of the vertical coupler as a function of the taper length. The inset shows the fundamental TE mode field transfer from the top GaAs waveguide to the beneath 4H-SiC waveguide.

To fabricate the hybrid quantum photonic chip, we first employ ion slicing technique to prepare wafer-scale 4H-SiCOI material. The process flow started with hydrogen ion implantation to a 4H-SiC wafer (**Figure 1**a), followed by a thermal annealing process to detach the 4H-SiC thin-film from its virgin substrate (**Figure 1**b). Thereafter, a direct wafer bonding technique was used to transfer the thin-film 4H-SiC onto a $SiO_2$/Si substrate to form a 4H-SiCOI material platform (**Figure 1**c). The fabricated 4H-SiCOI has 4-inch wafer size with average thickness of about 300 nm and surface roughness of 0.5 nm (**Figure 1**d). More details regarding the fabrication methods of the 4H-SiCOI material can be found in our previous work[29]. Next, nanofabrication processes including electron beam lithography and reactive ion etching were employed to fabricate photonic devices on the 4H-SiCOI material platform (see **Supporting Information**). To generate single photons on the photonic chip, we employed InGaAs self-assembled quantum dots (QDs) as source of single photons, and they were epitaxially grown in the middle of a 180 nm GaAs nanomembrane supported by a 200 nm thick $Al_{0.75}Ga_{0.25}As$ sacrificial layer on a (001) GaAs substrate. Nanophotonic waveguides were patterned on GaAs nanomembrane as schematically shown in **Figure 1**e. The $Al_{0.75}Ga_{0.25}As$ sacrificial layer was removed by a wet chemical etching process (**Figure 1**f). We then transfer the free-standing GaAs waveguides to a 4H-SiC photonic chip using a PDMS stamp assisted pick-and-place technique we previously developed[50-51]. With the help of a high-resolution microscopy, we can deterministically transfer GaAs nanophotonic waveguides onto the 4H-SiC chip with a precision of tens of nanometers. To enable efficient light transfer from QDs to the 4H-SiC photonic chip, a bilayer vertical coupler consisting of a GaAs tapered waveguide (400 nm wide) on top of a 4H-SiC ridge waveguide (1.0 μm wide and 150 nm thick) is designed, as illustrated by the sketch in **Figure 1**g. The GaAs waveguide has a 6 μm long taper which ensures an efficient coupling of QDs emission to the 4H-SiC waveguide. **Figure 1**h displays the simulated coupling efficiency of the bilayer vertical coupler. It shows that up to 98% of the fundamental transverse electric field like (TE-like) mode light can be coupled to the 4H-SiC waveguide for the given taper length. The inset displays a cross-sectional image of the fundamental TE-like mode propagation in the coupler. To evaluate the overall coupling efficiency of the QD emission to the 4H-SiC waveguide, the QD-to-waveguide coupling efficiency, the so-called $\beta$-factor[52], is calculated by employing the FDTD simulation method. A single QD emitter acting as an in-plane dipole source at 900 nm was modeled in the GaAs waveguide. It is found that about 51.5% of QD radiation can be coupled to the fundamental TE-like mode of the GaAs waveguide, and this suggests an overall coupling efficiency of about 50.4% for the QD emission to the 4H-SiC waveguide.

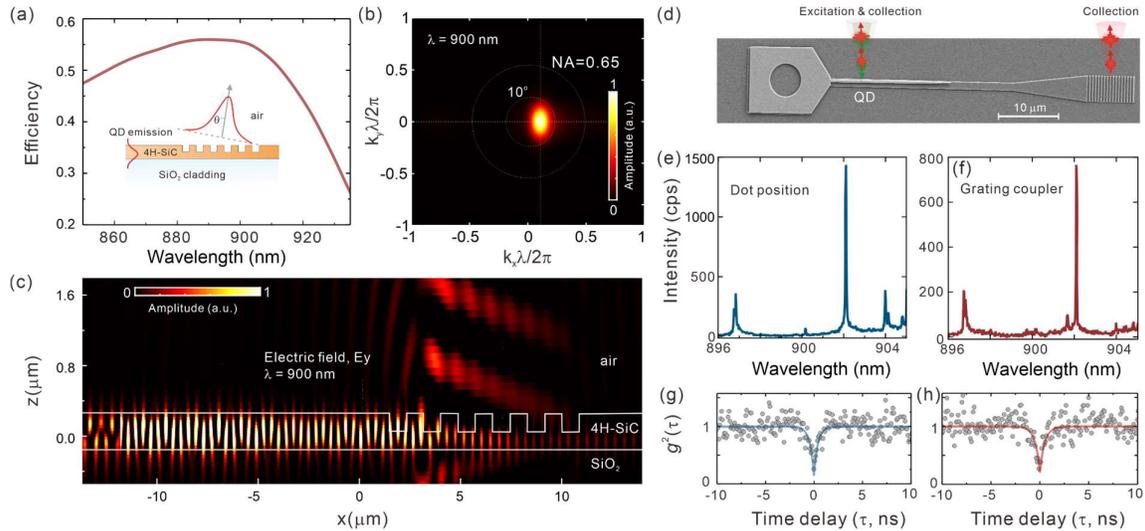

**Figure 2:** a) Simulated coupling efficiency of the focusing grating coupler. b) Far field pattern of the grating coupler at the wavelength of 900 nm. The inner and outer circles indicate the numerical aperture of the objective lens and the solid angle of 10° respectively. c) Cross-sectional image of the fundamental TE-like mode field scattered vertically by the focusing grating coupler. d) SEM image of the fabricated straight 4H-SiC waveguide. The left end of the waveguide is integrated with a GaAs tapered waveguide for single-photon generation while the other end is terminated with a focus grating coupler. e) and f) PL spectra of dotA collected from the source position and the grating coupler respectively. g) and h) Normalized second-order correlation function for the dominant spectral lime in e) and f).

To interface the quantum photonic chip and external free-space measurement apparatus, a focus grating coupler is designed to extract the light from the 4H-SiC waveguide. The grating coupler consists of an array of shallow-etched teeth with a period of 472 nm and a duty cycle of 50%. The inset in **Figure 2**a illustrates the cross-sectional profile of the focusing grating coupler. The mode propagating along the 4H-SiC waveguide is diffracted upwards by the grating teeth with an emission angle of $\theta$. **Figure 2**a shows the calculated coupling efficiency of the grating from which an optimized coupling efficiency of about 55.7% is obtained for the fundamental TE-like mode at 900 nm. The far-field plot in **Figure 2**b reveals a good directionality of the outgoing beam with an emission angle $\theta \sim 10°$, within the range of the numerical aperture of our objective lens (N.A.= 0.65). **Figure 2**c shows the main electrical field component of the fundamental TE-like mode ($E_y$). We can see that most of the TE-like mode light can be scattered vertically into free space, while a small fraction of the light is scattered downward, which accounts for the main source of loss of the grating coupler. It should be noted that this loss could be significantly suppressed by employing chirped teeth geometry[53]. Furthermore, we note that the grating offers a moderate coupling efficiency in a spectral range of about 10 nm, and this enables an effective light coupling over the wavelength range of our QDs ensemble.

With the above design of active and passive photonic components required for implementation of hybrid quantum photonic chip, we now present single-photon generation and routing in the chip. To this end, we fabricated a straight 4H-SiC waveguide as shown by the SEM image in **Figure 2**d. By employing an unbalance waveguide geometry, the propagation loss of our 4H-SiC waveguide is measured to be 7.5 dB/cm (see details in **Supporting Information**). A QDs-containing tapered GaAs waveguide was integrated on the left end of the 4H-SiC waveguide while the other end is terminated with a focusing grating coupler. The length of the waveguide is about 40 μm and the integrated tapered GaAs waveguide is about 20 μm. In order to verify the successful coupling of QDs emission embedded inside the GaAs nanophotonic waveguides to the beneath 4H-SiC waveguide, the fabricated chip together with the abovementioned straight waveguides was mounted in a closed-cycle cryostat and cooled to about 6 K. QDs in the tapered waveguide are excited with a 532 nm continuous wave (CW) laser through a microscope objective lens with NA=0.65. The micro-photoluminescence (µPL) is collected with the same objective and coupled into a single-mode fiber, which enables spatial filtering, thus only collecting the light scattered out by a specific grating. The PL signal collected by the optical fiber is then dispersed by a spectrometer with a focus length of 750 mm and then directed into a thermoelectrically cooled Si charge-coupled detector (CCD) array (see the **Supporting Information** for details of the experimental setup). With this experimental arrangement, we can separately collect photon emission from the QD source position and the grating coupler at the same time. **Figure 3**e and **3**f show the PL spectra of a single QD (denoted as dotA) embedded inside the tapered waveguide. By comparing the two spectra collected from the source position and the grating coupler, identical spectral features including the peak positions and line widths have been observed, which suggests a successful generation and routing of photons along the straight 4H-SiC waveguide. Subsequently, we characterize the photon emission properties by measuring the second-order photon auto-correlation function ($G^{(2)}(\tau)$). This measurement is implemented by selecting the dominant emission peak at 902.3 nm (1.3742 eV). The normalized auto-correlation function ($g^{(2)}(\tau)$) histograms are presented in the insets of **Figure 3**g and **3**h. Suppressed dips at the zero-time delay ($\tau$=0) are clearly seen, indicating an anti-bunching behavior for the photon emission. We then fit the anti-bunching curve with a theoretical model $g^{(2)}(\tau)=1-(1-g^{(2)}(0))\exp(-|\tau|/\tau_0)$, and the auto-correlation functions at the zero-time delay $g^{(2)}(0)$ are then fitted to be 0.26±0.06 and 0.22±0.03. These values are well below than the classical limit of 0.5, indicating a sub-Poissonian photon statistics for the photon emission from the chip-integrated single QD. It should be noted that the non-zero values of $g^{(2)}(0)$ contribute a multi-photon emission probability, which is likely from the contribution of unwanted photons emitted from the vicinity of the QD due to the above-band excitation and partially form the dark counts of the silicon avalanche photodiodes (about 100 cps)[54]. The purity of the single photons could be significantly improved by using a resonate excitation scheme[55-56] and high-performance superconducting nanowire single-photon detectors[57-58] with higher detection efficiency and much lower dark counts.

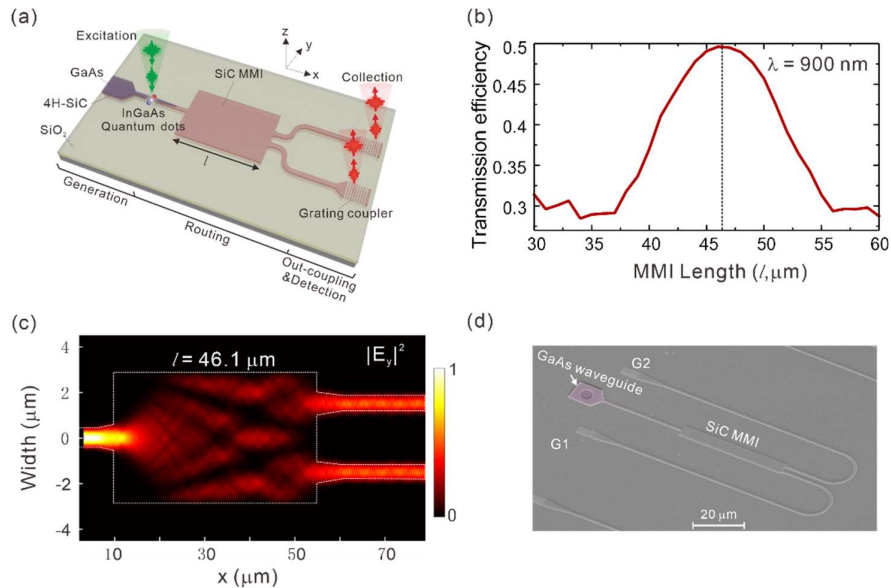

**Figure 3**: a) Sketch of the 1×2 MMI device designed on the 4H-SiCOI material platform. A tapered GaAs nanophotonic waveguide is integrated on the input waveguide of the MMI device. The two output ports are terminated with focusing grating coupler. b) Simulated transmission efficiency of one output port as a function of the MMI device length. c) Electric field intensity ($|E_y|^2$) of the MMI device for the symmetric transmission efficiency. d) SEM image of the fabricated MMI device. The purple color indicates the tapered GaAs nanophotonic waveguide transferred on the input waveguide of the MMI device.

To further evaluate the potential of our hybrid quantum photonic chip, we functionalize the photonic chip by realizing an on-chip beamsplitter operation with deterministic single photons. Indeed, on-chip beamsplitter operation represents a core implementation for the realization of photonic quantum logic gates and other advanced quantum functionalities in the framework of linear photonic quantum computing[59]. To achieve this goal, a multimode interference (MMI) device is designed and fabricated. Compared to the commonly used directional coupler, MMI devices feature with self-imaging effect by which the input light field can be reproduced in single or multiple images at outports with different power ratios[60-61]. For a proof-of-concept demonstration, the simplest MMI device, *i.e.*, 1×2 MMI device, is designed in our work (**Figure 3**a). The width of the MMI section is chosen to be 6 μm for supporting multi-mode at the wavelength of 900 nm. By sweeping the length of the MMI device ($l$) in the FDTD simulations, we find that symmetric transmission efficiency of 49.6% at each output port occurs at $l = 46.1$ μm for the desired wavelength (**Figure 3**b). The electric field profile plotted in **Figure 3**c visualizes this symmetric transmission efficiency. As the field is lunched into the input waveguide, we can see that the fundamental mode filed couples to higher-order modes of the MMI device and subsequently they interfere with each other. As a result, a pair of field patterns whose amplitudes equal to $1/\sqrt{2}$ of the initial field are observed, which forms a chip-integrated 50:50 beamsplitter. According to the simulations, we fabricated a MMI device as the SEM image shows in **Figure 3**d. On top of the input waveguide, a QDs-containing GaAs tapered waveguide is integrated for photon generation (the purple color). Two grating couplers are terminated at both output ports of the device for simultaneously collecting photon emission from the self-assembled QDs.

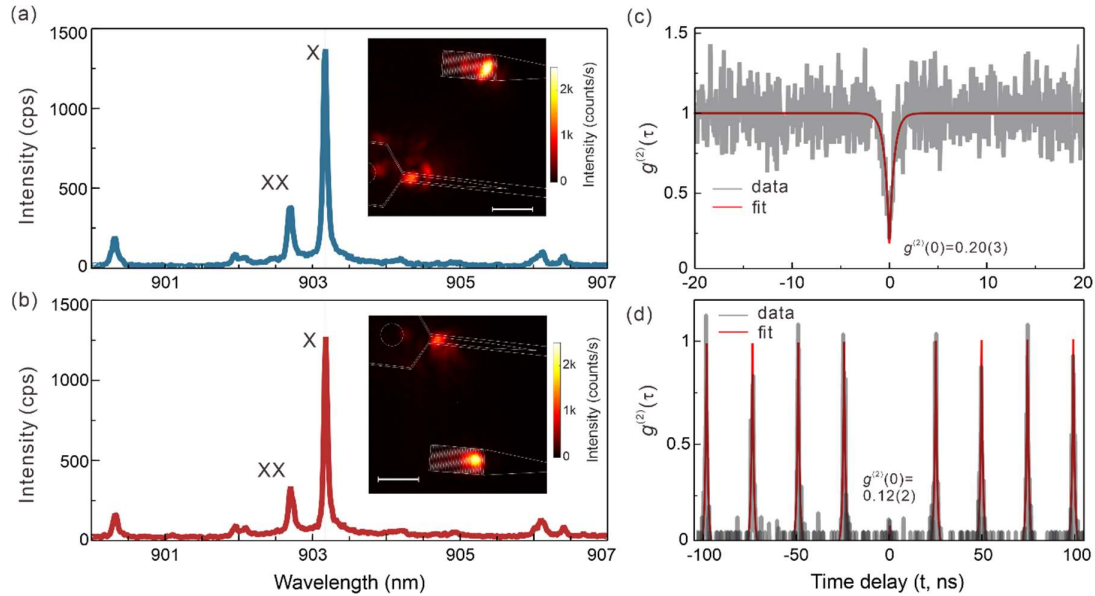

**Figure 4:** a) and b) PL spectra of dotB and they are collected from the top and the bottom grating coupler respectively. The insets show PL maps of the MMI device, where the scalebar is 8 μm in length. Normalized cross-correlation functions of the exciton photons collected from the separated grating couplers. The measurements are conducted under c) CW laser excitation and d) pulsed laser excitation.

**Figure 4**b) and **4**c show the PL spectra of a single QD inside the tapered GaAs waveguide (denoted as dotB) and they are collected from the two separated out-coupling grating couplers. Both of the spectra show the same excitonic emission peaks including the exciton (X) and biexciton (XX) emission lines. To see how the photons emitted by dotB are mapped in the 1×2 MMI device and scattered by the out-coupling grating coupler, we experimentally carry out two-dimensional raster scans of the PL signal for the device. Since the device footprint (120×50 μm$^2$) exceeds the scanning range of our experimental setup (~20 × 20 μm$^2$), we separately map the two distant grating couplers whilst keeping the QD excited with a CW laser. As shown by the insets of **Figure 4**b and **4**c, aside from the bright spot at the QD source position, we can clearly see two bright spots located at the grating couplers. These PL maps unambiguously confirm a successful routing of QD emission from the tapered GaAs waveguide to the grating coupler through the 1×2 MMI device. Thereafter, we send the PL signal at the output ports of the MMI device to two single-photon avalanche diodes (SPADs) and carry out a chip-based Hanbury-Brown-Twiss (HBT) measurement. In combination with external spectrometers, we spectrally select the X emission line and perform cross-correlation measurements. **Figure 4**c shows the normalized second-order correlation function for the X photons under 532 nm CW laser excitation. Clear anti-bunching behavior with $g^{(2)}(0) = 0.20 \pm 0.03$ has been observed. **Figure 4**d shows the normalized correlation function $g^{(2)}(\tau)$ of the X emission line for pulsed laser excitation at 532 nm. The periodic peaks together with the absent peak at zero-time delay reveal triggered single-photon emission events[62]. We fit the histogram in **Figure 4**d and determine $g^{(2)}(0) = 0.12 \pm 0.02$. Moreover, the time separations between the neighboring auto-correlation peaks are found to be 25 ns, which is consistent with the repetition rate (40 MHz) of the pulsed laser. With these results, we have successfully validated a chip-based beamsplitter operation with triggered single photons emitted by a single self-assembled QD. This implementation may constitute a major step toward advanced quantum functionalities in the 4H-SiC hybrid quantum photonic chip.

In summary, we have experimentally demonstrated heterogenous integration of self-assembled QDs based SPSs with a novel 4H-SiCOI photonic platform. A PDMS stamp assisted pick-and-place method is used to deterministically integrate the QDs-containing GaAs photonic waveguides onto the 4H-SiC photonic waveguides, thus forming a hybrid quantum photonic chip. We have fabricated straight 4H-SiC ridge waveguides and realized on-chip generation and routing of single-photon states. Especially, chip-based beam splitter operation has been achieved by designing and fabricating a 1×2 MMI device with a symmetric power splitting ratio. By combining off-chip SPADs, we have successfully performed a chip-based HBT measurement by utilizing the MMI-based beamsplitter. The successful realization of deterministic single-photon generation and routing in the 4H-SiC based hybrid quantum photonic chip provides a viable route toward CMOS-compatible and fast reconfigurable integrated quantum photonic circuit with deterministic SPSs.

**Supporting Information**

Supporting Information is available from the Wiley Online Library or from the author.

**Acknowledgements**

This work has been supported by National Key RD Program of China (2017YFE0131300), Science and Technology Commission of Shanghai Municipality (16ZR1442600, 20JC1416200), Shanghai Municipal Science and Technology Major Project (2017SHZDZX03), Shanghai Rising-Star Program (19QA1410600), Program of Shanghai Academic/Technology Research Leader (19XD1404600), National Natural Science Foundation of China (No. 12074400, U1732268, 61874128, 61851406, and 11705262), Frontier Science Key Program of Chinese Academy of Sciences (No. QYZDY-SSW-JSC032).

Yifan Zhu, Wenqi Wei and Ailun Yi contributed equally to this work.

Received: ((will be filled in by the editorial staff))
Revised: ((will be filled in by the editorial staff))
Published online: ((will be filled in by the editorial staff))